\documentclass[aps,amsmath,twocolumn,a4paper,jcp]{revtex4}
\usepackage{graphicx}
\usepackage[usenames]{color}
\usepackage{amssymb}
\usepackage{amsmath}
\usepackage{amsfonts}
\usepackage{mathrsfs}

\renewcommand{\epsilon}{\varepsilon}

\begin{document}

\title{The chain sucker: translocation dynamics of a polymer chain into a
long narrow channel driven by longitudinal flow}

\author{Kaifu Luo}
\altaffiliation{Author to whom the correspondence should be addressed}
\email{kluo@ustc.edu.cn}
\affiliation{CAS Key Laboratory of Soft Matter Chemistry, Department of
Polymer Science and Engineering, University of Science and Technology of
China, Hefei, Anhui Province 230026, P. R. China}
\author{Ralf Metzler}
\email{metz@ph.tum.de}
\affiliation{Physics Department, Technical University of Munich,
85747 Garching, Germany}
\affiliation{Physics Department, Tampere University of Technology,
FI-33101 Tampere, Finland}

\date{\today}

\begin{abstract}
Using analytical techniques and Langevin dynamics simulations, we investigate
the dynamics of polymer translocation into a narrow channel of width $R$ embedded
in two dimensions, driven by a force proportional to the number of monomers in the channel.
Such a setup mimics typical experimental situations in nano/micro-fluidics.
During the the translocation process if the monomers in the channel can
sufficiently quickly assume steady state motion, we observe the scaling
$\tau\sim N/F$ of the translocation time $\tau$ with the driving force $F$ per
bead and the number $N$ of monomers per chain. With smaller channel width $R$,
steady state motion cannot be achieved,
effecting a non-universal dependence of $\tau$ on $N$ and $F$.
From the simulations we also deduce the waiting time distributions under various
conditions for the single segment passage through the channel entrance.
For different chain lengths but the same driving
force, the curves of the waiting time as a function of the translocation coordinate
$s$ feature a maximum located at identical
$s_{\mathrm{max}}$, while with increasing the driving force or the channel width
the value of $s_{\mathrm{max}}$ decreases.

\end{abstract}

\pacs{87.15.A-, 87.15.H-}

\maketitle

\section{Introduction}

The transit of biopolymeric chains across membranes through
nanopores, the \emph{translocation\/} process, is a recurrent theme in
cell biology \cite{alberts,muthu,meller,rapoport}. In biological membranes
the nanopore is typically constituted by a channel protein, the transmembrane
proteins of the haemolysin family being a prominent example \cite{griffiths}.
One family member, $\alpha$-haemolysin, is also
often employed in single nanopore setups \emph{in vitro\/} \cite{nanopores}.
More recently, solid state nanopores in artificial supports such as silicon
compound membranes are manufactured by ion or electron beam techniques and
open up the possibility for controlled technological applications
\cite{solidstate,beam}.

In biology, important examples for translocation processes are the passage
of RNA or proteins through pores in the nuclear membrane, of proteins across
mitochondrial and chloroplast membranes or through the endoplasmatic reticulum,
the cell-to-cell exchange of DNA across the cell walls, or the viral injection
of RNA and DNA molecules \cite{alberts}. In technology, biopolymer translocation
is envisaged to be useful for rapid DNA sequencing, gene therapy, and,
ultimately, towards controlled drug delivery.

The transport of biopolymers through a nanopore has attracted broad interest
in the statistical physics community, as it represents a challenging problem
in polymer physics \cite{Sung,Muthukumar99,MuthuKumar03,Chuang,Luo1,Luo4,Dubbeldam1,
Kantor,Luo2,Matysiak,Dubbeldam2,Panja,Luo3,Aniket,Slater,Sakaue,Milchev}.
A quantity of particular interest is the average
translocation time $\tau$ as a function of the chain length $N$, usually
assumed to follow a scaling law $\tau\sim N^{\alpha}$. The scaling exponent
$\alpha$ hereby reflects the efficiency of the translocation process. Thus,
for completely directed (ratcheted) motion, one would expect $\tau\sim N$,
while for normally diffusive unbiased translocation $\tau\sim N^2$ \cite{Muthukumar99,Sung}.
Generally, the value of $\alpha$ differs from these limiting behaviors, as effected by the
entropic degrees of freedom of the chain-to-be-translocated. In particular,
when the translocation dynamics is governed by unbiased
anomalous diffusion \cite{report}
of the form $\langle s^2(t)\rangle\sim t^{\beta}$ ($0<\beta<1$) in terms of
the translocation coordinate $s$, the value of the scaling exponent $\alpha=2/
\beta$ becomes larger than 2 \cite{Kantor,mekla}.

During the passage of a long flexible chain through a nanopore, its two
extremities cannot access the volume opposite the membrane, and the monomer(s)
in the pore are more of less unable to move. This reduction of the accessible
degrees of freedom of the polymer involves a considerable entropic barrier
during the pore passage. While unbiased translocation has been argued to
feature the same scaling exponent $\alpha=1+2\nu$ as the free diffusion of the
polymer by its radius of gyration \cite{Chuang,Luo4,Luo1}, where the Flory exponent
is $\nu=0.588$ in 3D and $\nu_{2D}=0.75$ in 2D \cite{deGennes,Doi,Rubinstein},
the associated prefactor is much larger \cite{Chuang}.
Thus efficient polymer translocation requires the presence of
driving forces. Most experimental \cite{nanopores,solidstate,beam} and
theoretical \cite{Kantor,Luo2,Matysiak,Dubbeldam2,Panja,Luo3,Aniket,Slater,Sakaue,Santtu}
studies focus on the driving force provided by an external applied electric field,
which mainly falls off across the pore. In this case, $\alpha$ depends on the translocation
dynamics. For slow translocation, i.e.,
under low driving force and/or high friction, $\alpha\approx1+\nu\approx
1.588$ in 3D \cite{Kantor,Luo8}, while $\alpha\approx 1.37$ in 3D \cite{Luo8}
for fast translocation due to the highly deformed chain conformation on the
trans side, reflecting a pronounced non-equilibrium situation. In addition,
for translocation achieved by pulling one end of the polymer with a constant
force, $\alpha\approx2.0$ was established \cite{Kantor,Huopaniemi2}.
The above examples of translocation involve constant forces throughout
the process. A different situation is encountered for translocating chains
whose passage is effected by binding proteins, so-called chaperones or
chaperonins \cite{Metz}. Due to the fact that the chaperones bind more than one monomer
of the chain-to-be-translocated, interesting variations occur in the effective
driving force due to the ``parking lot effect''.

\begin{figure}
\includegraphics[width=8.2cm]{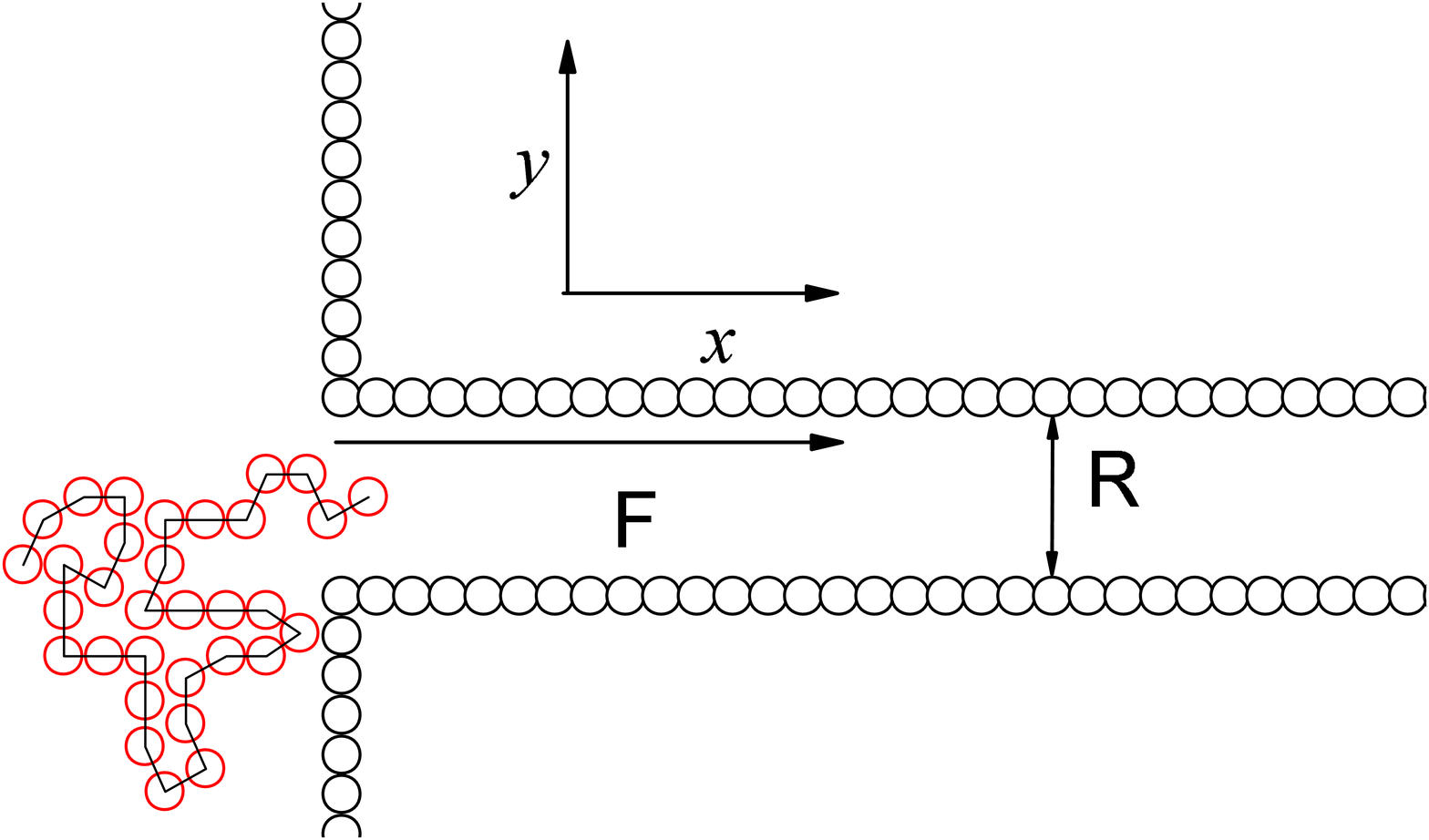}
\caption{Schematic representation of polymer translocation into a long 2D
channel of width $R$. The driving force acting on the bead in the channel
can be provided by an applied intra-channel electric field or a uniform flow
of the solvent in the channel. We assume that the net force on the chain is
proportional to the number of monomers, $s(t)$, in the chain.}
\label{Fig1}
\end{figure}

Here we consider a case motivated by state-of-the-art fluidic nanochannel setups,
in which the driving force grows constantly with the progress of the translocation
process, until a saturation is reached when the entire chain is in the channel.
In such setups one or multiple nanochannels branch off a micron-sized feeder
channel (access hole) \cite{thamdrup,tegenfeldt}.
The driving force is provided by the liquid flow: while
the flow field is quite low in the feeder channel, the flow velocity is much
higher inside the narrow channel. Once one end of the chain enters the
nanochannel, it is veritably sucked inside, the overall drag force on the chain
increasing as more and more of the monomers enter the channel. Similarly, one
might think of setups in which an electric field falls off in a long channel.
If each monomer of the chain-to-be-translocated carries a net charge, the force
acting on the chain will increase with the proportion of the chain inside the
channel.
This is not the case for the relatively short channel \cite{Tobias}, where the
driving force cannot be proportional to the translocated monomers once the first
monomer has passed through the pore.
In many cases such fluidic setups are built in a pseudo two-dimensional
geometry, i.e.,
the channel is sandwiched between two parallel walls, that are close to each
other. Having such scenarios in mind we consider the geometry depicted in
Fig.~\ref{Fig1}. We investigate this problem using analytical techniques and
Langevin dynamics (LD) simulations. The paper is organized as follows. In
Section II, we describe our model and the simulations technique. In Section
III, we present and discuss our results. Finally, the Conclusions are drawn
in Section IV.

\section{Model and methods}
\label{chap-model}

In our numerical simulations, the polymer chains are modeled as bead-spring
chains of Lennard-Jones (LJ) particles with the Finite Extension Nonlinear
Elastic (FENE) potential \cite{Kremer}. Excluded volume interactions between
beads are taken into consideration by a short range repulsive LJ potential
\begin{equation}
U_{\mathrm{LJ}}(r)=4\epsilon\left[{\left(\frac{\sigma}{r}\right)}^{12}-
{\left(\frac{\sigma}{r}\right)}^6\right]+\epsilon,
\end{equation}
for $r\le 2^{1/6}\sigma$ and 0 for $r>2^{1/6}\sigma$. Here, $\sigma$ is
the diameter of a bead and $\epsilon$ is the potential depth. The
connectivity between neighboring beads is modeled as a FENE spring with
\begin{equation}
U_{\mathrm{FENE}}(r)=-\frac{1}{2}kR_0^2\ln\left(1-\frac{r^2}{R_0^2}\right),
\end{equation}
where $r$ is the distance between consecutive beads, $k$ is the spring constant,
and $R_0$ is the maximum allowed separation between connected beads.

We consider the geometry shown in Fig. \ref{Fig1}: two strips with separation
$R$ consisting of stationary particles with a bead-bead distance $\sigma$ form
the walls of the channel, as well as the ``membrane'' containing the pore.
Between all bead-wall particle pairs, there exists the same short range
repulsive LJ interaction as described above. In the LD simulations, each bead
is subjected to conservative, frictional, and random forces, respectively,
with \cite{Allen}
\begin{equation}
m\mathbf{\ddot{r}}_i=-\mathbf{\nabla}({U}_{\mathrm{LJ}}+{U}_{\mathrm{FENE}})+
\mathbf{F}_{\mathrm{ext}}-\xi\mathbf{v}_i+\mathbf{F}_i^R,
\end{equation}
where $m$ is the bead mass, $\xi$ is the friction coefficient for a single
bead, $\mathbf{v}_i=\mathbf{\dot{r}}_i$ is the bead velocity,
and $\mathbf{F}_i^R$ is the random
force satisfying the fluctuation-dissipation theorem. The external force is
expressed as $\mathbf{F}_{\mathrm{ext}}=F\hat{x}$, where $F$ is the external
force strength per bead, exerted on the translocating beads located inside
the channel, and $\hat{x}$
is a unit vector in the direction along the channel. Therefore, the driving
force for the whole chain is proportional to the number of translocated beads.
Experimentally, this external driving force acting on the beads in the channel
can be provided by an applied intra-channel electric field or a uniform flow
of the solvent in the channel.

In the present work, we use the LJ parameters $\epsilon$ and $\sigma$ and the
bead mass $m$ to fix the energy, length and mass scales respectively. The time
scale is then given by $t_{\mathrm{LJ}}=(m\sigma^2/\epsilon)^{1/2}$. The
dimensionless parameters in our simulations are $R_0=2$, $k=7$, $\xi=0.7$
and $k_{B}T=1.2$ unless otherwise stated. The Langevin equation is integrated
in time by a method described by Ermak and Buckholz \cite{Ermak} in 2D.

Initially, the first bead of the chain is placed just inside the channel
(at $x=0.75$, $y=0$), while the remaining beads are under thermal
collisions described by the Langevin thermostat to obtain an equilibrium
configuration. The translocation time is defined as the time duration between
the beginning of the translocation and the last monomer entering into the
channel. Note that, in contrast to many translocation setups, we allow the
chain to escape from the channel back towards the cis side of the membrane.
Typically, we average our data over 1000 independent runs.

Here, we should mention that the effect of hydrodynamic interactions is
neglected in our simulation. Recent lattice Boltzmann \cite{Pablo,Fyta}
and molecular dynamics \cite{Slater} simulation results show that hydrodynamics
is screened out in a narrow pore, which is the case here and in the experiments.

\section{Results and discussion}
\label{chap-results}

In this Section, we present our simulation results and discuss them in view
of scaling arguments provided in the following Subsection.

\subsection{Scaling arguments}

In standard translocation models, where the driving force acts solely in
a short channel, only a single bead currently inside the channel experiences
the driving force. For sufficiently slow translocation in the short nanopore,
i.e., under low driving force and/or high friction, the exponent $\alpha$ in
the scaling law $\tau\sim N^{\alpha}$ for the translocation time $\tau$ can
be estimated by the balance of driving force $F$ and frictional force. The
velocity of the center of mass along the direction of the driving force is
$v\sim F/\xi$. The chain is not severely deformed during slow translocation
processes, and the chain moves a distance of $R_g$ during the translocation.
Thus, the translocation time becomes $\tau\approx R_g/v\sim N^{1+\nu}\xi/F$
\cite{Kantor,Luo2,Luo8}. Similarly, for polymer translocation under a constant pulling
force acting on one \emph{chain end}, the polymer travels a distance of
$R_{\parallel}\sim N$ during translocation, and $\tau\approx R_{\parallel}/v
\sim N^2\xi/F$ \cite{Huopaniemi2}. However, if the external driving force is
a function of time as for the present problem in a long channel with an
intra-channel force, the translocation dynamics becomes different, as outlined
in the following.

For a polymer of chain length $N$ in a good bulk solvent in two dimensions (2D),
the radius of gyration of the chain $R_g$ scales as $R_g\sim N^{\nu_{2D}}
\sigma$, where $\nu_{2D}=0.75$ is the Flory exponent in 2D, and $\sigma$ is the
segment length. For a polymer confined between two strips embedded in 2D, the
chain will extend along the channel to form blobs of size $R$, as long as $R>
\sigma$. Each blob contains $g\sim(R/\sigma)^{1/\nu_{2D}}$ beads, and the
number of blobs is $n_b=N/g\sim N(\sigma/R)^{1/\nu_{2D}}$. The free
energy cost for the chain confinement in units of $k_BT$ is $\mathcal{F}=N
(\sigma/R)^{1/\nu_{2D}}$ \cite{deGennes}. The blob picture then predicts the
longitudinal size of the chain to be $R_{\parallel}=n_bR\sim N\sigma(
\sigma/R)^{1/\nu_{2D}-1}\sim NR^{-1/3}$ \cite{deGennes}. The
longitudinal relaxation time $\tau_{\parallel}$ is defined as the time needed
by a polymer to move a distance of the order of its longitudinal size, $R_{
\parallel}$. Thus, $\tau_{\parallel}$ scales as $\tau_{\parallel}\sim R_{
\parallel}^2/\widetilde{D}\sim N^{3}R^{-2/3}$, where ${\widetilde{D}}\sim
1/N$ is the diffusion constant of the chain.

For a completely confined polymer in a 2D long channel under an external
field, the total driving force is $NF$ with $F$ being the force acting on
one bead. If the chain moves a distance of the order of $R_{\parallel}$ under
weak driving forces, the time cost can be estimated as
$\tau_{\mathrm{ss}}\sim R_{\parallel}/v \sim NR^{-1/3}/F$,
where $v\sim F$ is the velocity of the polymer.
Under strong driving forces, $R_{\parallel}\sim N$ is still correct \cite{Luo10}, but
$R_{\parallel}\sim R^{-1/3}$ breaks down due to the deformation of the blob and the
elongation of the chain under the driving force.
Thus, one can only obtain $\tau_{\mathrm{ss}} \sim N/F$.
For polymer translocation through a long channel with length $L\gg R_g$, we
therefore obtain $\tau_{\mathrm{ss}}\sim L/v\sim 1/F$, which is
\emph{independent\/} of $N$ \cite{Yong}. This \emph{steady state\/} behavior in the
channel, valid sufficiently far away from the two channel ends, is due to
the fact that the force $F$ acts uniformly on each monomer. The behavior is
different when we consider the translocation time $\tau$ measuring the time
from threading the first monomer into the channel until the moment when the
entire chain is inside the channel.

\begin{figure}
\includegraphics[width=8.2cm]{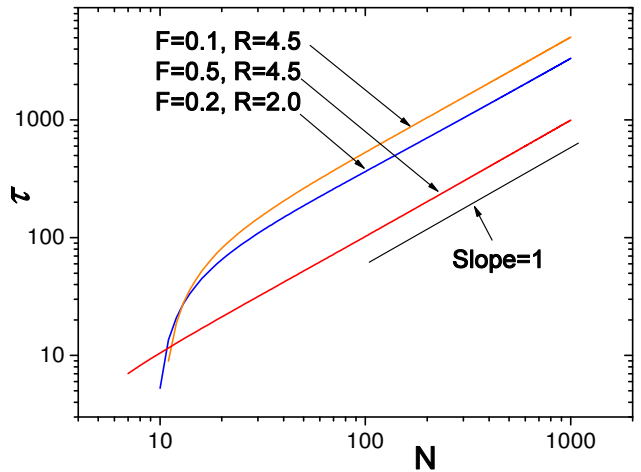}
\caption{Translocation time $\tau$ predicted in Eq. (\ref{eq3}) is plotted as function of
chain length $N$ for different channel widths $R$ and driving forces $F$. }
\label{Fig2}
\end{figure}

To assess the capture of a polymer into the long channel under the influence of
the intra-channel field, the situation is somewhat complicated due to the fact
that the driving forces are proportional to the number of translocated beads.
However, we still can give a rough estimate, if the chain monomers sufficiently
quickly reach a steady state in the channel.
If $s(t)$ monomers are located inside the channel, the chain
on the $cis$ side experiences the friction force $F_{cis,f}$ and the entropic
force $F_{cis,e}$, while the translocating part experiences the friction force
$F_{trans,f}$, the entropic force $F_{trans,e}$ and the driving force $s(t)F$.
According to the balance of driving force, frictional forces and entropic
forces, we obtain the force relation
\begin{equation}
F_{cis,f}+F_{cis,e}+F_{trans,f}+F_{trans,e}=s(t)F.
\end{equation}
Based on the blob picture,
\begin{equation}
\label{eq1}
F_{trans,f}\approx\xi s(t)\frac{dR_{\parallel}(t)}{dt}\approx CR^{-1/3}\xi s(t)\frac{ds(t)
}{dt},
\end{equation}
with $C$ being a constant. For a chain with size $N=128$ in a channel with $R=4.5$ in
units of $\sigma$, the longitudinal size is $R_{\parallel}\approx90.73$, and then we
obtain $C=1.17$.
Once more than $g$ monomers enter into the channel, $F_{trans,e}\approx 2.12
k_BT/R$ according to field-theoretical methods for 2D geometries
(strips) and hard-wall boundaries \cite{Burkhardt}.
Under the assumption that the chain on the $cis$ side is close to its equilibrium state
and is not severely deformed ($F_{cis,e}R_g(N-s(t))<k_BT$), $F_{cis,e}$ can be estimated as
$F_{cis,e}\approx k_BT L_x/[R_g^2(N-s(t))]$ \cite{deGennes}, with $L_x$ being the elongation of the chain
along the $x$ direction.
%
In addition, $F_{cis,f}=\xi\sum_{i=s(t)+1}^{N}v_i$, which under
sufficiently slow translocation is negligible compared with $F_{trans,f}$, due
to the much slower velocity. Thus, at the beginning of the translocation,
$F_{cis,e}$ plays an important role. However, once more than $g$ monomers
enter the nanopore, $F_{cis,e}$ can be neglected compared with $F_{trans,e}$.
Due to the fact that the driving force $s(t)F$ increases over time, after
reaching a critical $s(t)$ the translocation dynamics is dominated by the
balance between the driving force $s(t)F$ and $F_{trans,f}$, particularly for
larger $F$ where $F_{trans,e}$ becomes negligible. Then the translocation
dynamics is dominated by
\begin{equation}
CR^{-1/3}\xi s(t)\frac{ds(t)}{dt}=s(t)F,
\end{equation}
such that
\begin{equation}
\label{eq2}
\tau\sim\xi\frac{N}{F}R^{-1/3}.
\end{equation}
If $F_{trans,e}$ cannot be neglected, once more than $g$ monomers enter into
the channel, the translocation dynamics is controlled by
\begin{equation}
CR^{-1/3}\xi s(t)\frac{ds(t)}{dt}+F_{trans,e}=s(t)F,
\end{equation}
such that
\begin{equation}
\label{eq3}
\tau\approx CR^{-1/3}\xi\frac{N}{F}-\frac{CR^{-1/3}\xi F_{trans,e}}{F^2}[1-
\ln(FN-F_{trans,e})].
\end{equation}
The results in Eqs. (\ref{eq2}) and (\ref{eq3}) are based on Eq. (\ref{eq1}),
where the equilibrium value of $R_{\parallel}$ is used. As mentioned above,
under strong driving forces $R_{\parallel}\sim N$ is still correct, but
$R_{\parallel}\sim R^{-1/3}$ breaks down. Therefore, we do not expect the
scaling exponents of $\tau$ with $R$ predicted in Eqs. (\ref{eq2}) and
(\ref{eq3}) to be strictly correct. However, we can still use Eq. (\ref{eq3})
with $F_{trans,e}\approx 2.12k_BT/R$ for the equilibrium state to estimate the
scaling of $\tau$ with $N$. Fig.~\ref{Fig2} demonstrates that the
scaling $\tau\sim N$ for long chains is fulfilled, and $F_{trans,e}$ is thus
important only for short chains.

If during the the translocation process the monomers in the channel cannot
sufficiently quickly assume steady state motion, the translocation dynamics is
more complicated, as discussed in the following based on numerical results.

\subsection{simulation results}

We now proceed to present our numerical results and discuss them in the light
of above scaling arguments.

\subsubsection{Translocation probability as function of the driving force
$F$ and channel width $R$}

\begin{figure}
\includegraphics[width=8.2cm]{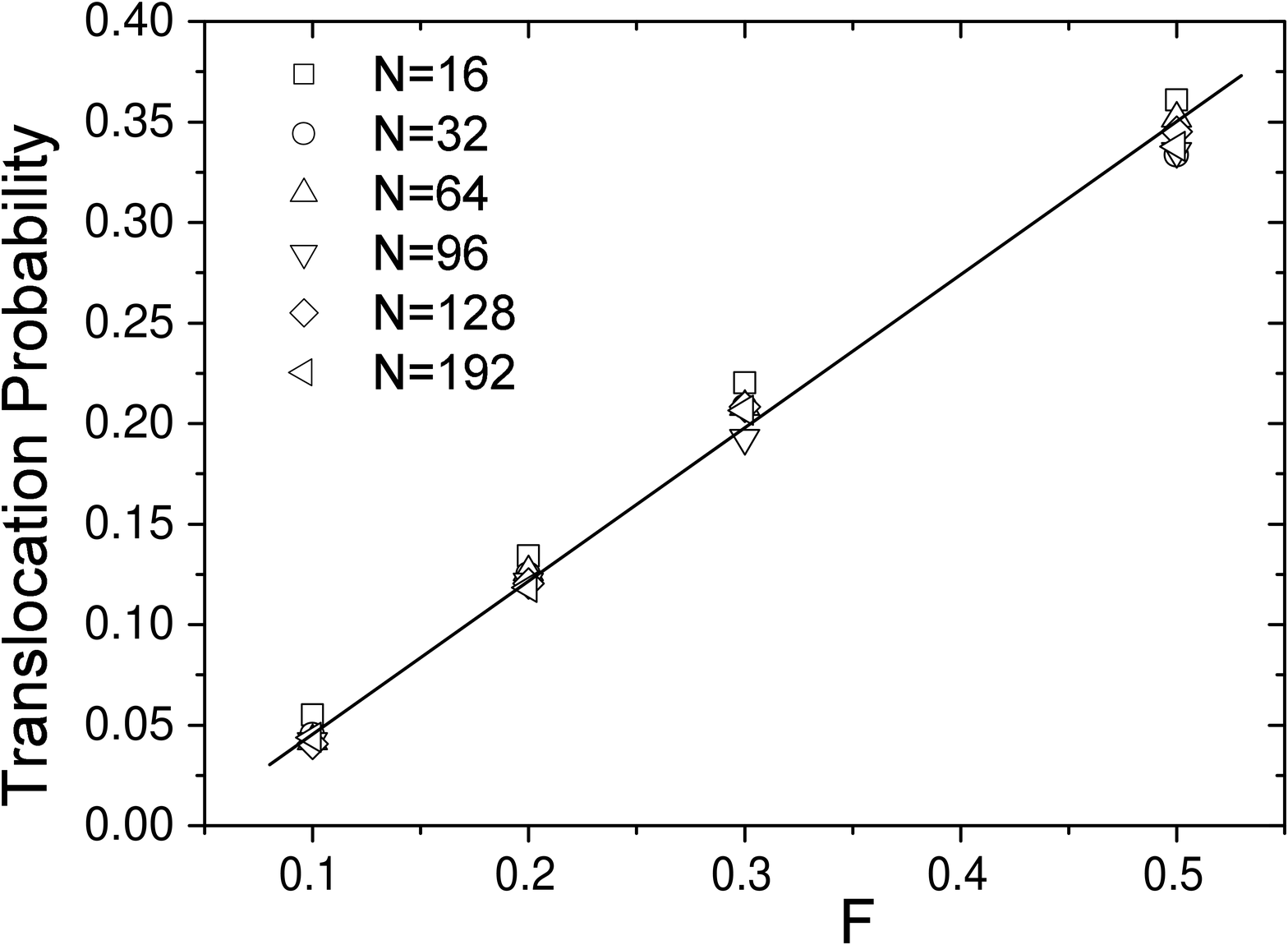}
\caption{Translocation probability as function of the driving force $F$ for
fixed channel width $R=4.5$. The results for various chain lengths approximately
coincide. The line indicates a linear relation between the translocation probability
and the driving force $F$ per beads.}
\label{Fig3}
\end{figure}

\begin{figure}
\includegraphics[width=8.2cm]{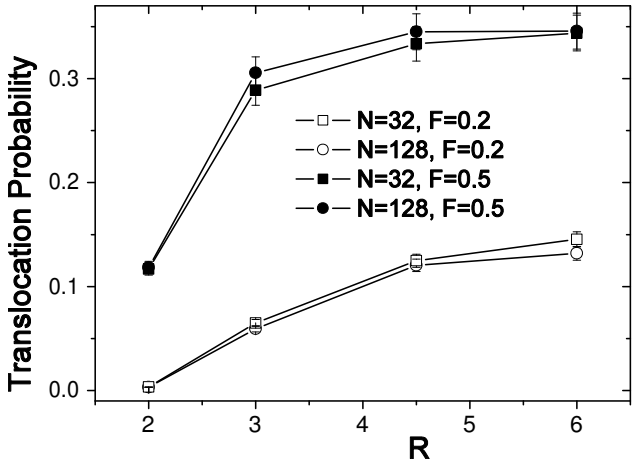}
\caption{Translocation probability as function of the channel width $R$ for
driving forces $F=0.2$ and 0.5, and for two chain lengths.}
\label{Fig4}
\end{figure}

In our simulations initially the first monomer is held in place right at the
channel entrance. After equilibration of the chain with fixed first monomer,
the constraint is relieved. At high driving force $F$, already acting on
the first monomer, the probability is relatively high that the chain will
actually completely move into the channel, and not escape back to the
cis side. At lower $F$ retraction becomes more frequent. In this context we
refer to a successful translocation as the event when the chain fully enters
the pore, i.e., all $N$ monomers are inside the channel such that the chain
reaches the translocation coordinate $s=N$. If the first monomer retracts
from the channel, the translocation is viewed unsuccessful. In Fig. \ref{Fig3}
we show the translocation probability of successful events as function of
the driving force $F$, and for different chain lengths $N$. For each value
of $F$ and $N$ we perform as many translocation attempts until 1000
successful translocation events are reached. In the investigated force range
the translocation probability increases almost linearly with increasing
translocation force $F$, and is approximately independent of the chain
length $N$. This implies that it is important to capture the first few
monomers in the channel, such that the overall force $s(t)F$ reaches a
sufficiently large value to prevent chain retraction from the pore.

Fig. \ref{Fig4} depicts the translocation probability as function of the
channel width $R$ for two chain lengths, at driving forces $F=0.2$ and 0.5. The
resulting curves show an increase from quite small channel width towards
wider channels, as expected. This is due to the decreased entropic
resistance against threading of the chain into the channel. The approximate
independence of the chain length suggests that this is, again, a more local
effect: it matters to succeed sucking the first few monomers into the
channel to ensure complete translocation. At larger values for $R$ we observe
a flattening of the curve, likely due to the elongation of the chain in the channel.

\subsubsection{Translocation times as function of the chain length $N$ and
channel width $R$}

\begin{figure}
\includegraphics[width=8.2cm]{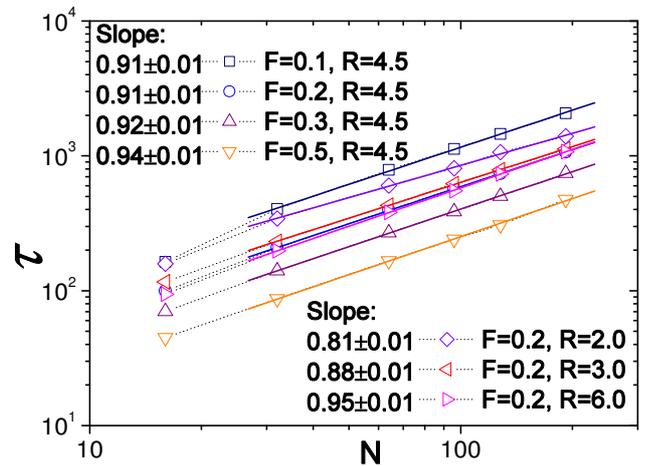}
\caption{Translocation time $\tau$ as function of chain length $N$ for
different channel widths $R$ and driving forces $F$.}
\label{Fig5}
\end{figure}

Fig. \ref{Fig5} shows the translocation time $\tau$ as function of the chain
length $N$ for different driving forces $F$ and channel widths $R$. For
$N\ge32$, a good scaling behavior $\tau\sim N^{\alpha}$ is observed, with
scaling exponent $\alpha\approx1$ for $F=0.1\ldots0.5$ and $R=4.5$, as well
as for $F=0.5$ and $R=6.0$. In these cases, the driving force dominates the
translocation dynamics, and the value of $\alpha$ is in agreement with the
result (\ref{eq2}) obtained from our scaling arguments. For $N<32$,
$F_{trans,e}\approx2.12k_BT/R$ increases and is no longer negligible for
weak driving forces as predicted in Eq. (\ref{eq3}) and plotted in Fig. \ref{Fig2}.
Decreasing $R$ for fixed $F=0.2$, the scaling exponent $\alpha$ is reduced to
$0.88\pm0.01$, and $0.81\pm0.01$ for $R=3.0$ and 2.0, respectively.
The relaxation time of the chain increases with decreasing $R$,
which leads to significant non-equilibrium situations. This factor leads to
a more complicated translocation dynamics,
that can no longer be grasped by our simple scaling arguments leading to
Eqs. (\ref{eq2}) and (\ref{eq3}). The resulting behavior, however, is good news: increasing
the overall force exerted on the chain during the process of threading the
chain into the channel leads to a more efficient translocation, with a
scaling exponent smaller than one. It should be noted that once the entire
chain is fully in the channel and reaches a steady state, the typical time
to cover a distance $R_{\parallel}$ becomes independent of $N$, as shown above.

\begin{figure}
\includegraphics[width=8.2cm]{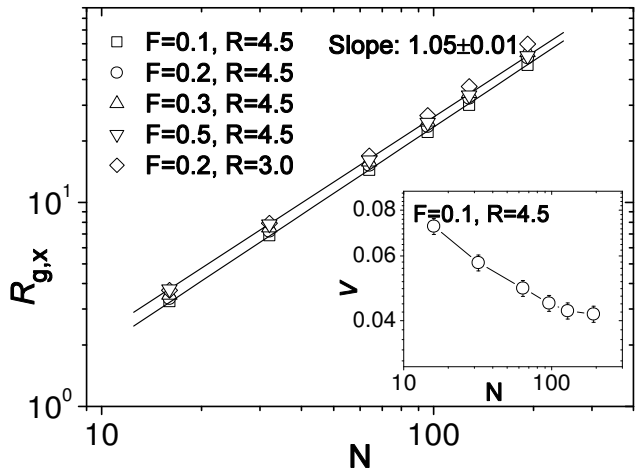}
\caption{Radius of gyration of the translocating polymer at the moment just
after translocation, i.e., full entrance into the channel, for different
translocation forces $F$ and channel widths $R$. The insert shows the translocation velocity $v$
as a function of the chain length $N$.}
\label{Fig6}
\end{figure}

\begin{figure}
\includegraphics[width=8.2cm]{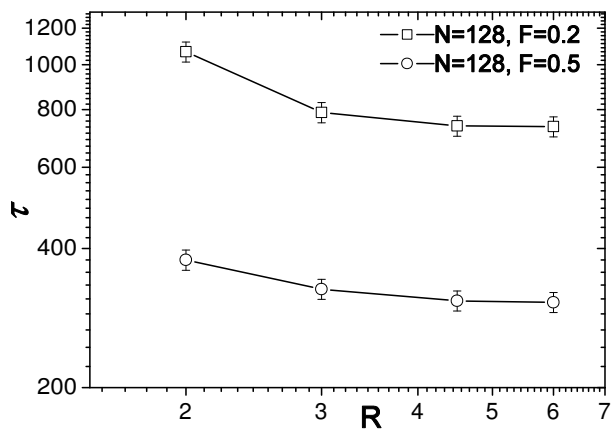}
\caption{Translocation time $\tau$ as function of the channel width $R$ for
$N=128$, and $F=0.2$ and 0.5.}
\label{Fig7}
\end{figure}

The radius of gyration of the polymer along the channel axis $x$, which is
proportional to $R_{\parallel}$, at the moment just after complete entrance
into the channel, for different values of the translocation force $F$ and
channel width $R$ turns out to follow the almost linear relation $R_{g,x}\sim
N$. This is demonstrated in Fig. \ref{Fig6}. This is perfectly in line with our
scaling arguments for $R_{\parallel}$ developed above. The inset in
Fig.~\ref{Fig6}
shows the translocation velocity, which decreases with chain length $N$, and
almost saturates for larger $N$.
With increasing channel width $R$, the translocation time decreases rapidly
and then saturates for larger $R$, as shown in Fig. \ref{Fig7}. Likely this
behavior is due to the decaying influence of the entropic force in the
channel, $F_{trans,e}\sim k_BT/R$ and the elongation of the chain under the driving force.

\subsubsection{Translocation time as function of the driving force $F$}

\begin{figure}
\includegraphics[width=8.2cm]{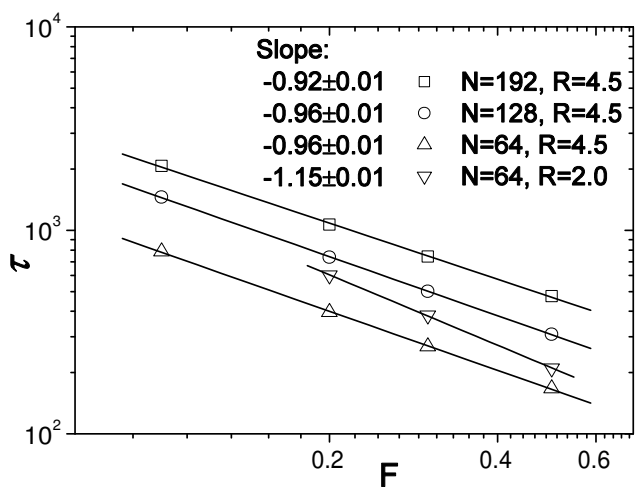}
\caption{Translocation time $\tau$ as function of the driving force $F$.}
\label{Fig8}
\end{figure}

Fig. \ref{Fig8} shows the translocation time $\tau$ as function of the driving
force $F$ for different chain lengths $N$. We observe a behavior close to the
predicted inverse proportionality with $F$, $\tau\sim F^{-1}$ from
Eq. (\ref{eq1}), for different chain lengths when the channel entropic force
$F_{trans,e}\sim k_BT/R$ is less pronounced.
However, for smaller $R$, such as $R=2.0$, the entropic force $F_{trans,e}$
plays a more important role in the translocation dynamics: thus, for $N=64$
and $R=2.0$, the chain cannot fully enter the nanopore at all, while for $F
\ge0.2$, we observe $\tau\sim F^{-1.15}$.

\subsubsection{Waiting time distribution}

\begin{figure}
\includegraphics[width=8.2cm]{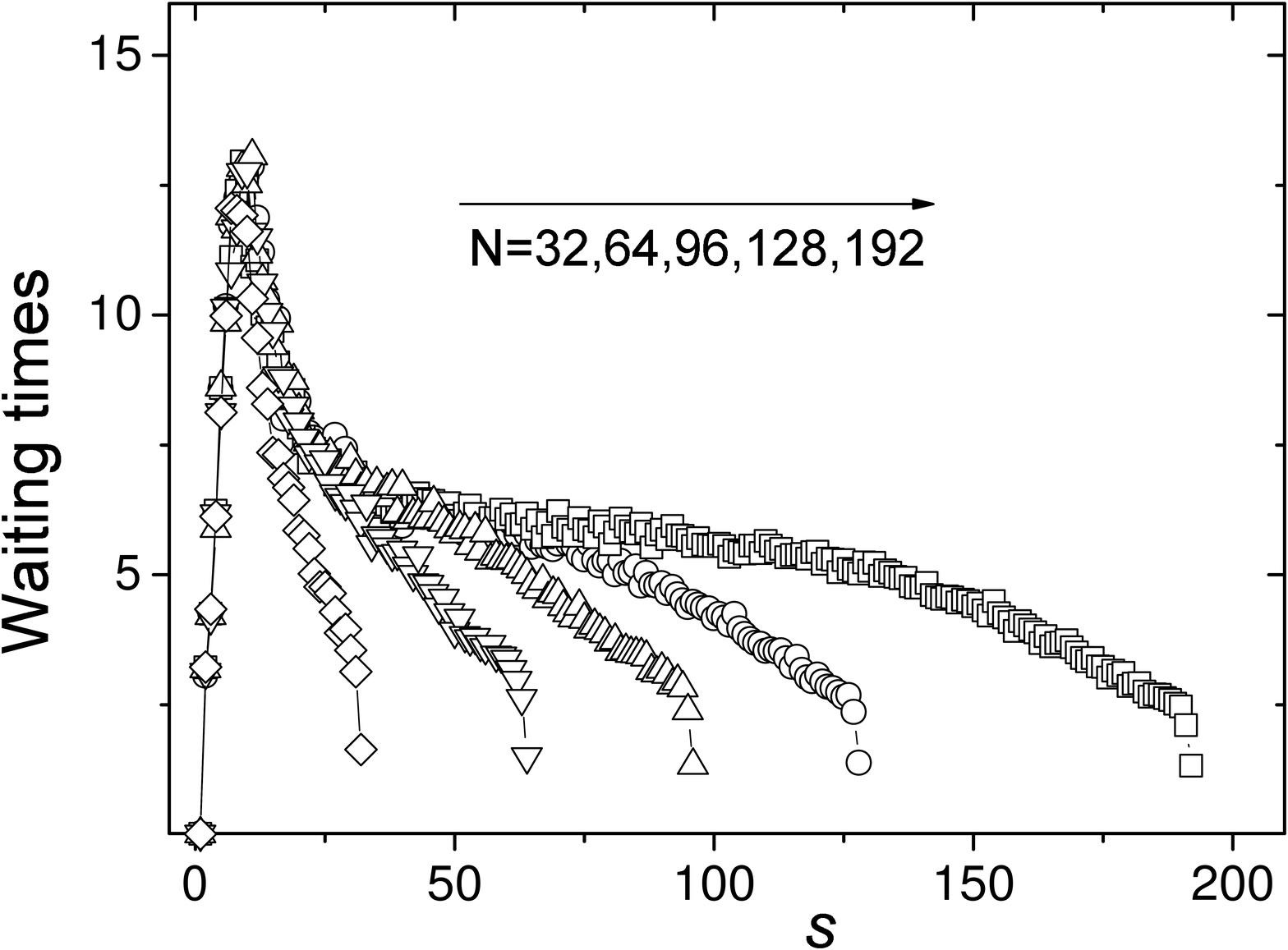}
\caption{Waiting time distribution for different chain lengths under driving
force $F=0.2$ and with channel width $R=4.5$. We define the waiting time of
bead $s$ as the average time between the events that bead $s$ and bead $s+1$
move away from the channel entrance.}
\label{Fig9}
\end{figure}

\begin{figure}
\includegraphics[width=8.2cm]{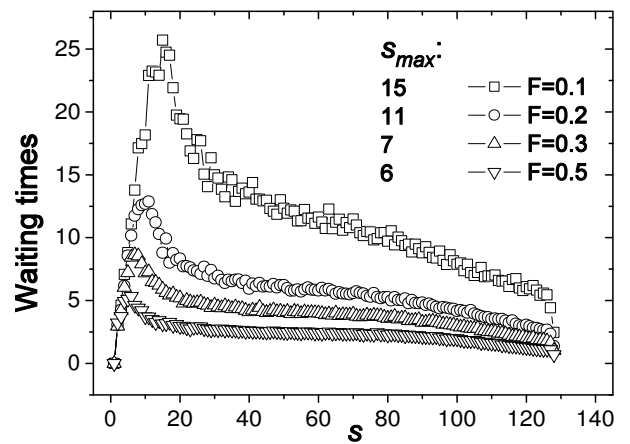}
\caption{Waiting time distribution under different driving forces for
chain length $N=128$ and channel width $R=4.5$.}
\label{Fig10}
\end{figure}

\begin{figure}
\includegraphics[width=8.2cm]{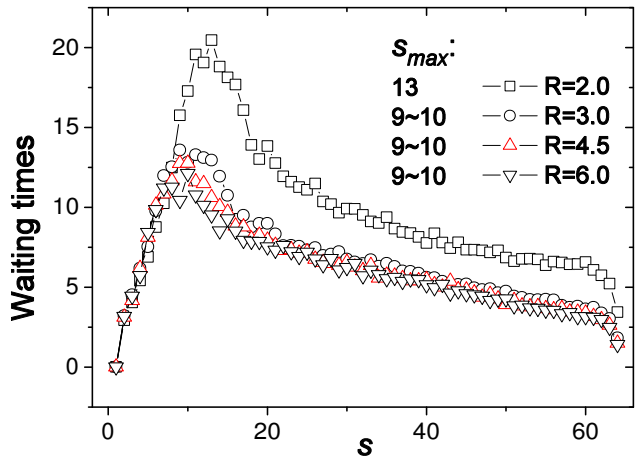}
\caption{Waiting time distribution for different channel width $R$
and chain length $N=64$ under driving force $F=0.2$.}
\label{Fig11}
\end{figure}

The dynamics of a single segment passing through the pore during translocation
is an important quantity considerably affected by different driving mechanisms.
The nonequilibrium nature of the translocation process has significant effects
on this quantity. We numerically calculated the waiting times for all beads in
a chain of length $N$. The waiting time of bead $s$ for successful translocation
hereby is defined as the average time between the events that bead $s$ and bead
$s+1$ move away from the channel entrance and further into the channel.

Fig. \ref{Fig9} shows the waiting time distribution for different chain lengths
under driving force $F=2.0$ and with channel width $R=4.5$. Interestingly, the
maximum waiting time occurs at $s_{\mathrm{max}}\approx 10$ for \emph{all measured\/} chain
lengths. With increasing $s$, the waiting time increases rapidly to the maximum
value, before a significantly softer decay. For increasing chain length a
plateau appears. This form of the waiting time distribution again suggests that
the initial capture of the chain in the channel is a local process, unaffected
by the remainder of the chain on the \emph{cis\/} side. At later stages of the
process the leftover monomers are sucked into the channel more easily, due to
the high force $s(t)F$ reached by the $s(t)$ monomers in the channel.
Fig. \ref{Fig10} shows the waiting time distribution under different driving
forces $F$, for chain length $N=128$ and channel width $R=4.5$. We observe that
the number of monomers corresponding to the maximum waiting time, $s_{\mathrm{
max}}$, shifts to smaller values with increasing driving force, as one would
expect. Increasing $R$, we find that $s_{\mathrm{max}}$ rapidly decreases but
saturates for larger $R$, compare Fig. \ref{Fig11}.

The coming into existence of non-equilibrium situations in the translocation
process, we display typical chain configurations in the Appendix.
Due to the increasing driving force $s(t)F$, the chain on the \emph{cis\/}
side experiences trumpet, stem-flower, and straight chain conformations
during the translocation process, see Figs.~\ref{Fig12}-\ref{Fig14}.

\section{Conclusions}
\label{chap-conclusions}

Using Langevin dynamics simulations, we investigated the dynamics of polymer
translocation into a long channel embedded in a two dimensional geometry,
mimicking
typical nano/micro-fluidic setups used, for instance, for DNA analysis. We
analyzed how the translocation dynamics depends on the chain length $N$ of
the chain-to-be-translocated, the driving force $F$, and the channel width
$R$.
During the the translocation process if the monomers in the channel can
sufficiently quickly assume steady state motion, we observe the scaling
$\tau\sim N/F$ of the translocation time $\tau$ with $F$ and $N$. With
smaller channel width $R$, steady state motion cannot be achieved,
effecting a non-universal dependence of $\tau$ on $N$ and $F$.
We also find that the waiting time distribution shows a maximum at the same
translocation coordinate $s_{\mathrm{max}}$ for different chain lengths, and
$s_{\mathrm{max}}$ decreases with increasing driving force or channel width.
We believe that this study opens up a new aspect to translocation studies, adding relevant
quantitative information on the growing field of biopolymer analysis in
nanochannels.

\begin{acknowledgments}
K. L. acknowledges support from the University of Science and Technology of
China through the CAS ``Bairen'' Program and the National Natural Science
Foundation of China (Grant No. 21074126). R. M. acknowledges support from
the Academy of Finland (FiDiPro scheme) and the Deutsche Forschungsgemeinschaft.
\end{acknowledgments}

\appendix*
\section{Translocation snapshots}

\begin{figure}
\includegraphics[width=8.2cm]{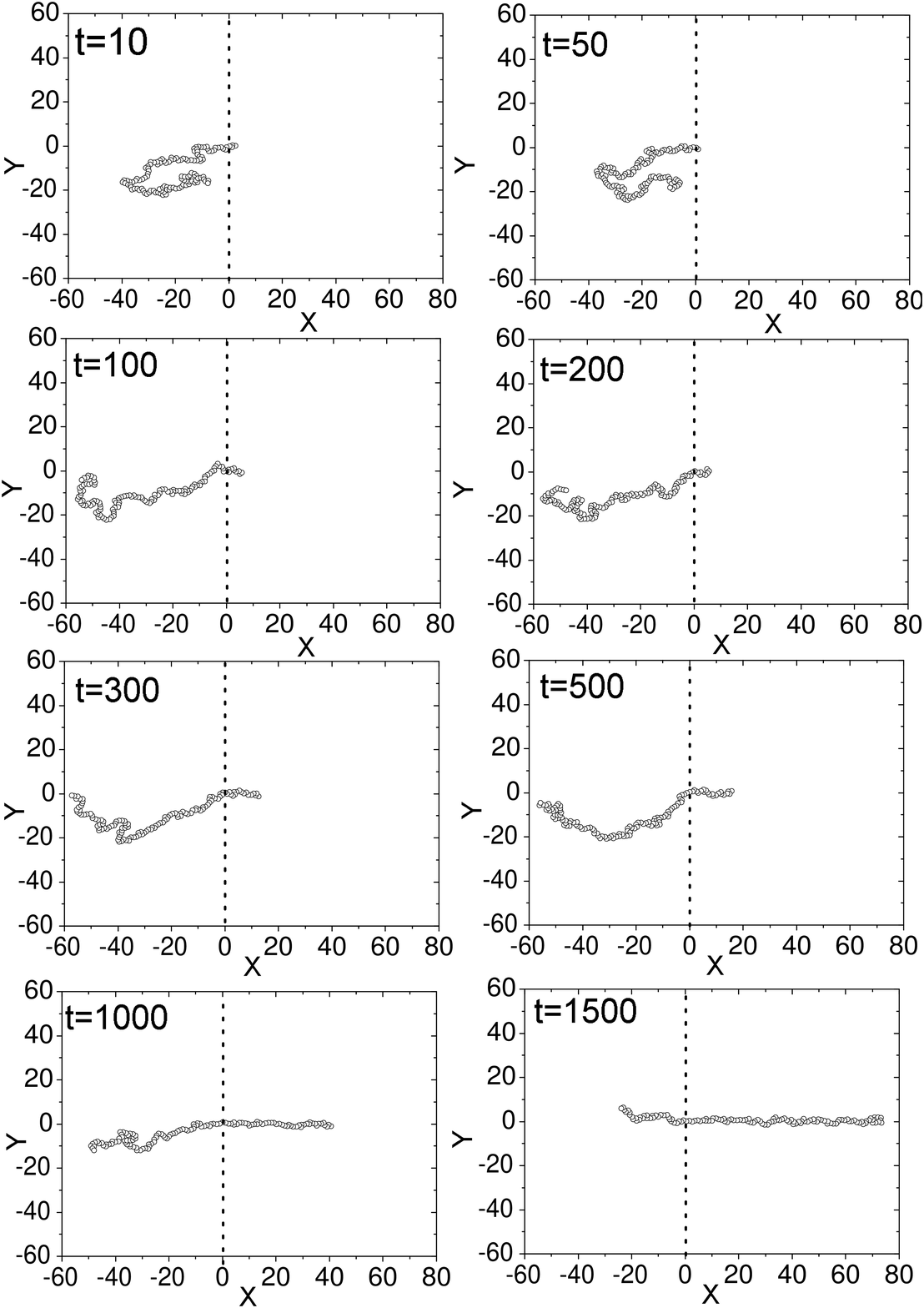}
\caption{Chain conformation during translocation for channel width $R=4.5$
and chain length $N=128$ under the driving force $F=0.1$.}
\label{Fig12}
\end{figure}

\begin{figure}
\includegraphics[width=8.2cm]{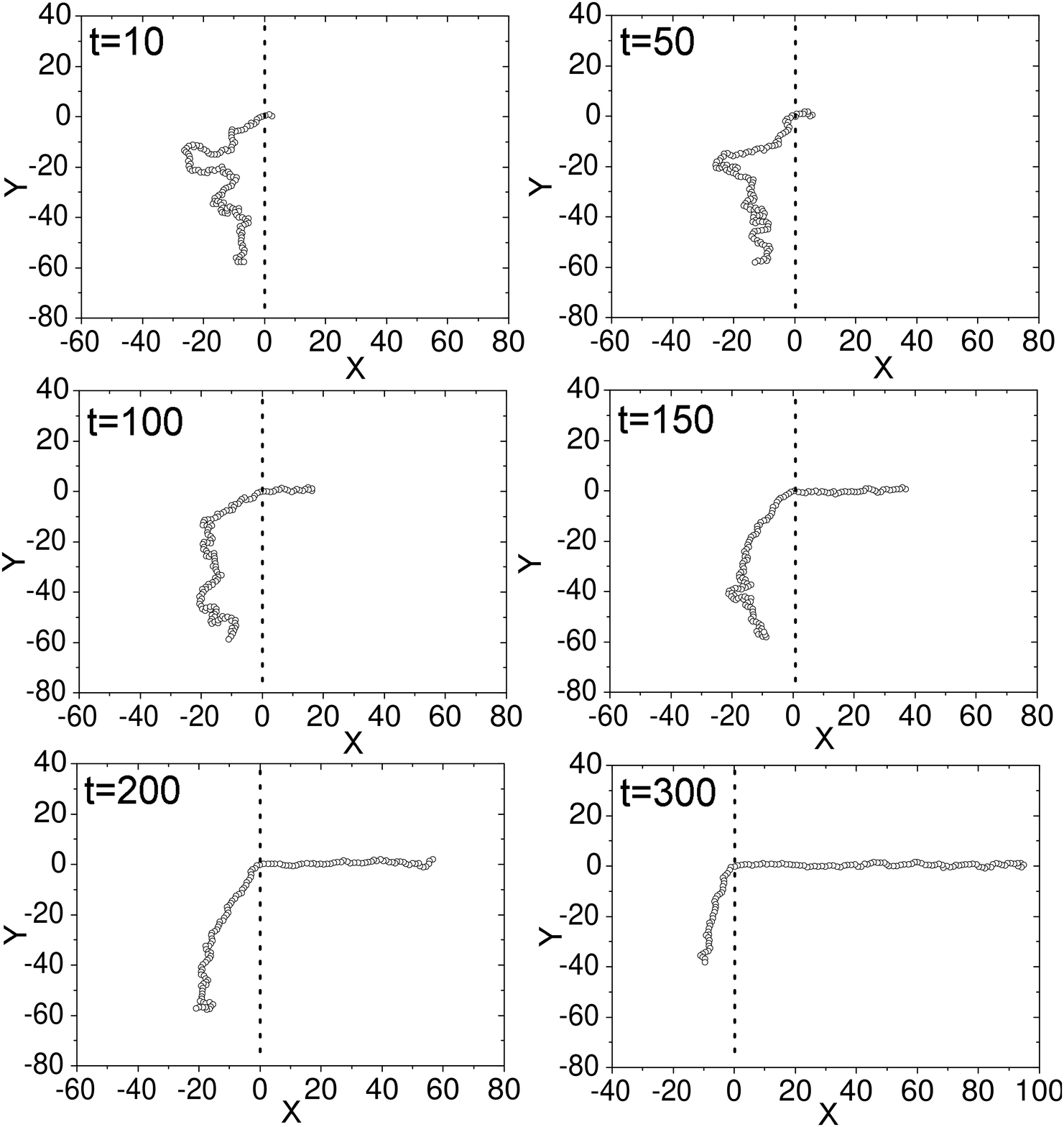}
\caption{Chain conformation during translocation for channel width $R=4.5$
and chain length $N=128$ under the driving force $F=0.5$.}
\label{Fig13}
\end{figure}

\begin{figure}
\includegraphics[width=8.2cm]{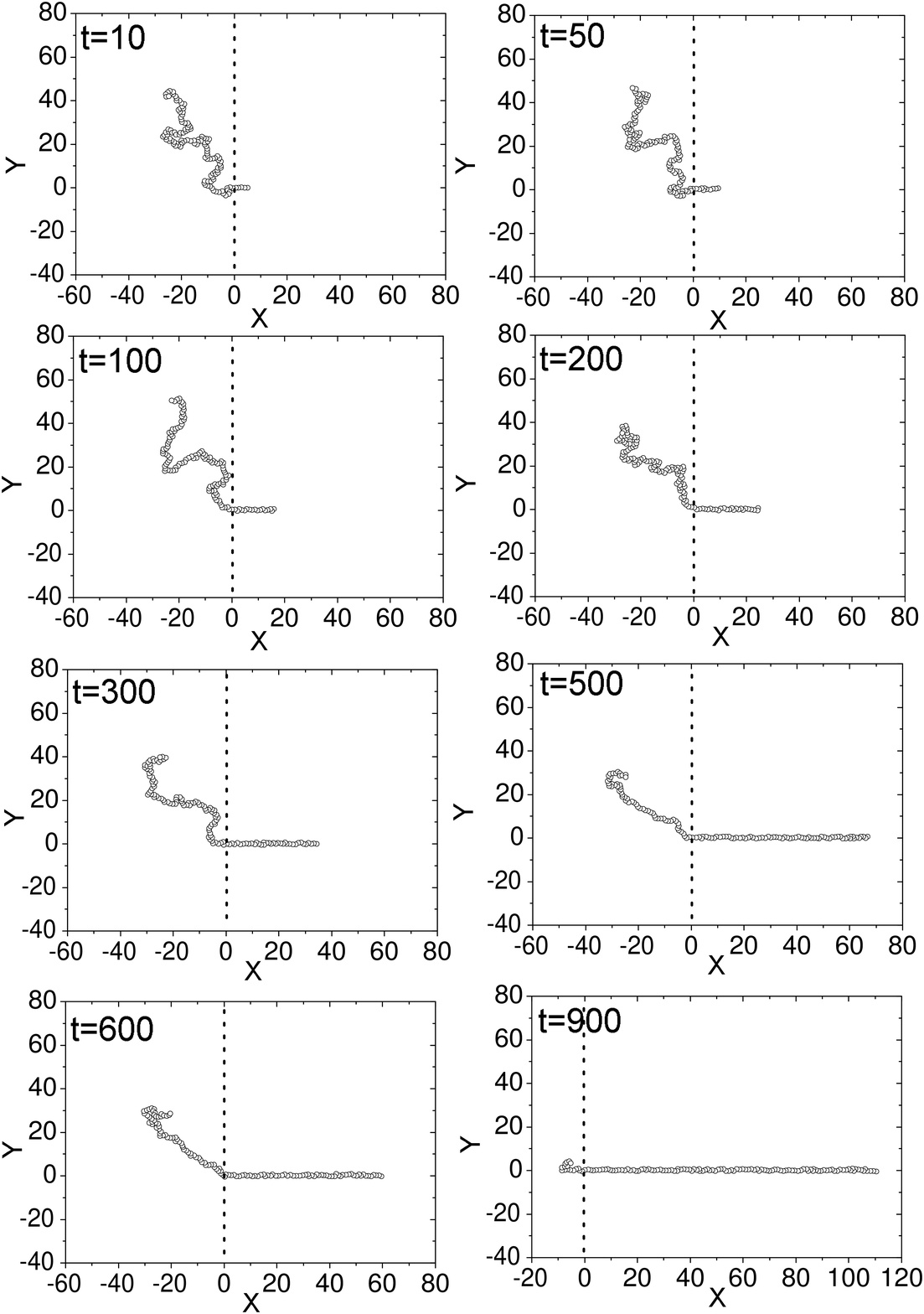}
\caption{Chain conformation during translocation for channel width $R=2.0$
and chain length $N=128$ under the driving force $F=0.2$.}
\label{Fig14}
\end{figure}

To get an idea of translocation process, we show the typical chain conformation for different stages of
the translocation process.  The chain length $N=128$.
Figs. \ref{Fig12} and \ref{Fig13} shows chain conformation during translocation for channel width $R=4.5$
under the driving force $F=0.1$ and $F=0.5$, respectively. For $F=0.1$, at $t=100$ the typical trumpet chain
conformation occurs. At $t=100$ and $t=1200$, we observe the typical stem-flower conformation. Finally, the
chain is almost straight at $t=1500$. For $F=0.5$, the chain is stretched to straight conformation at $t=300$.
Figs. \ref{Fig14} shows chain conformation during translocation for channel width $R=2.0$ under the driving
force $F=0.2$. The chain in the channel is very straight during the translocation.

\end{document}